\documentclass[multphys,vecphys]{svmult}

\usepackage{makeidx}         % allows index generation
\usepackage{graphicx}        % standard LaTeX graphics tool
\usepackage{multicol}        % used for the two-column index
\usepackage[bottom]{footmisc}% places footnotes at page bottom
\makeindex             % used for the subject index
\begin{document}

\title*{Intergal field spectroscopy survey of classical LBV stars in M33
}
\author{O. Sholukhova \inst{1},  P. Abolmasov\inst{1},
S. Fabrika \inst{1}, V. Afanasiev \inst{1}\and
M. Roth \inst{2}}
\institute{Special Astrophysical Observatory, Russia
\texttt{olga@sao.ru}
\and Astrophysikalisches Institut Potsdam, An der Sternwarte 16, D-14482,
Potsdam, Germany \texttt{mmroth@aip.de}}
\maketitle

\begin{abstract}
Five well-known LBV stars in M33 were observed
with the Multi-Pupil Fiber Spectrograph (MPFS) on the 6-m Russian telescope.
We observed LBVs  var\,A,  var\,B,  var\,C,  var\,2 and var\,83.
In three of them,
var\,2,  var\,83,  var\,B, large-scale nebulae were found with sizes
from 15 pc and larger. The nebula shapes are complex,  like one-side
tails or conical nebulae. They all are related to their LBV stars.
In var\,2 and var\,83 stars we found radial velocity gradients 15--30~km/s
across their nebulae.
The stars  var\,A and  var\,C do not show extended nebulae,
but nebular lines are certainty present in their spectra.
\end{abstract}

\section{Observations and data reduction}
\label{sec:1}

This observation were carried as continuation of our program of studing of
LBV-candidate stars in M33.
Here we presented results of observations on the Russian 6-m telescope
with the integral field spectrograph MPFS
(Afanasiev et al., 2001) in November 2004.
The integral field unit of 16$\times$16 square spatial elements
covers a region of 16"$\times$16" ~on the sky.
Integral field spectra were taken in
the spectral range 4000~-- 6800~\AA\AA\ with a seeing from 1.0
to 1.5" (FWHM). Data reduction was
made using procedures developed in IDL environment (version 6.0) by
V.\,Afanasiev, A.\,Moiseev and P.\,Abolmasov and include all the standard steps.

\begin{figure}
\includegraphics[width=5.7cm]{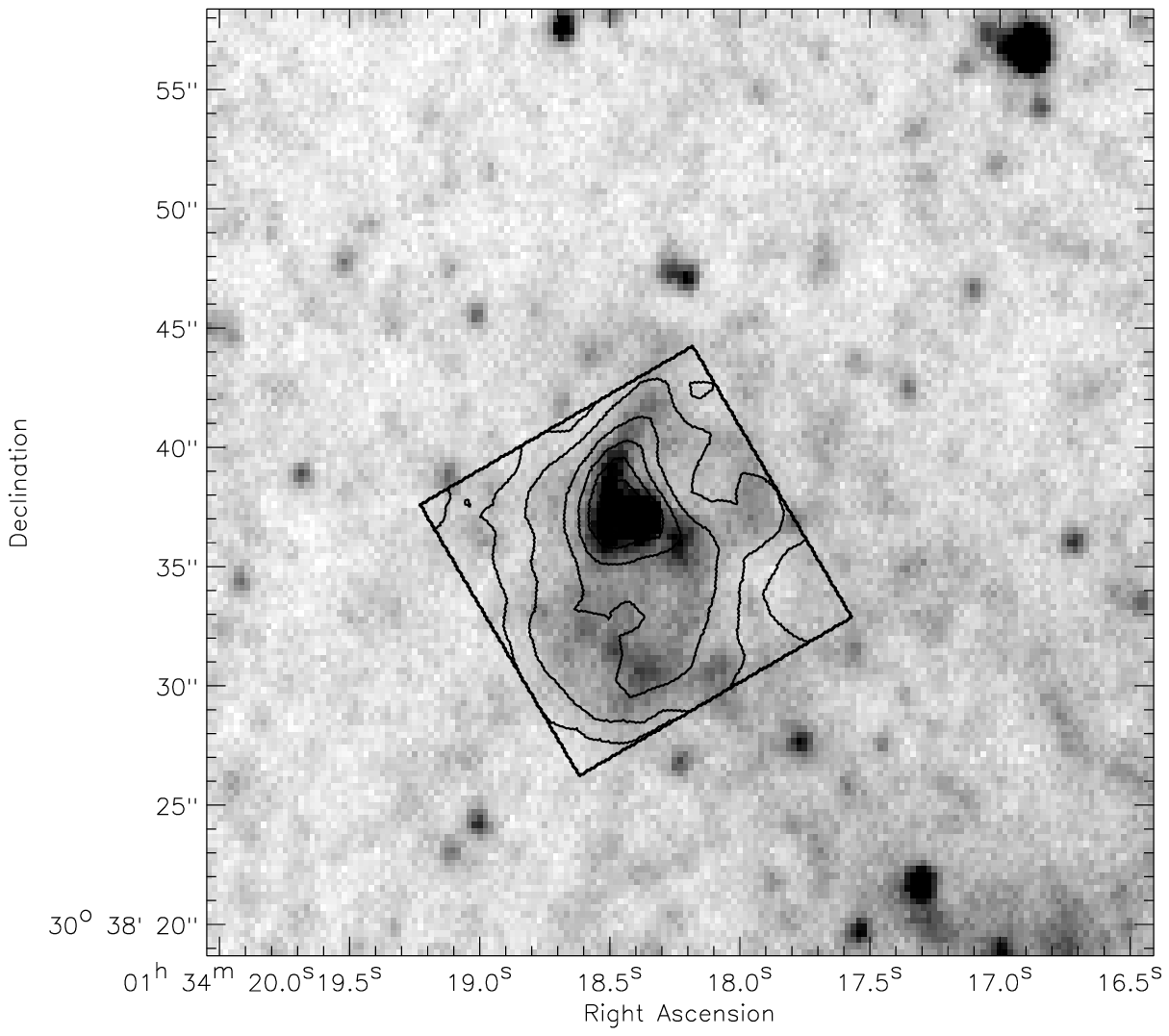}\hspace{0.5cm}
\includegraphics[width=4.8cm]{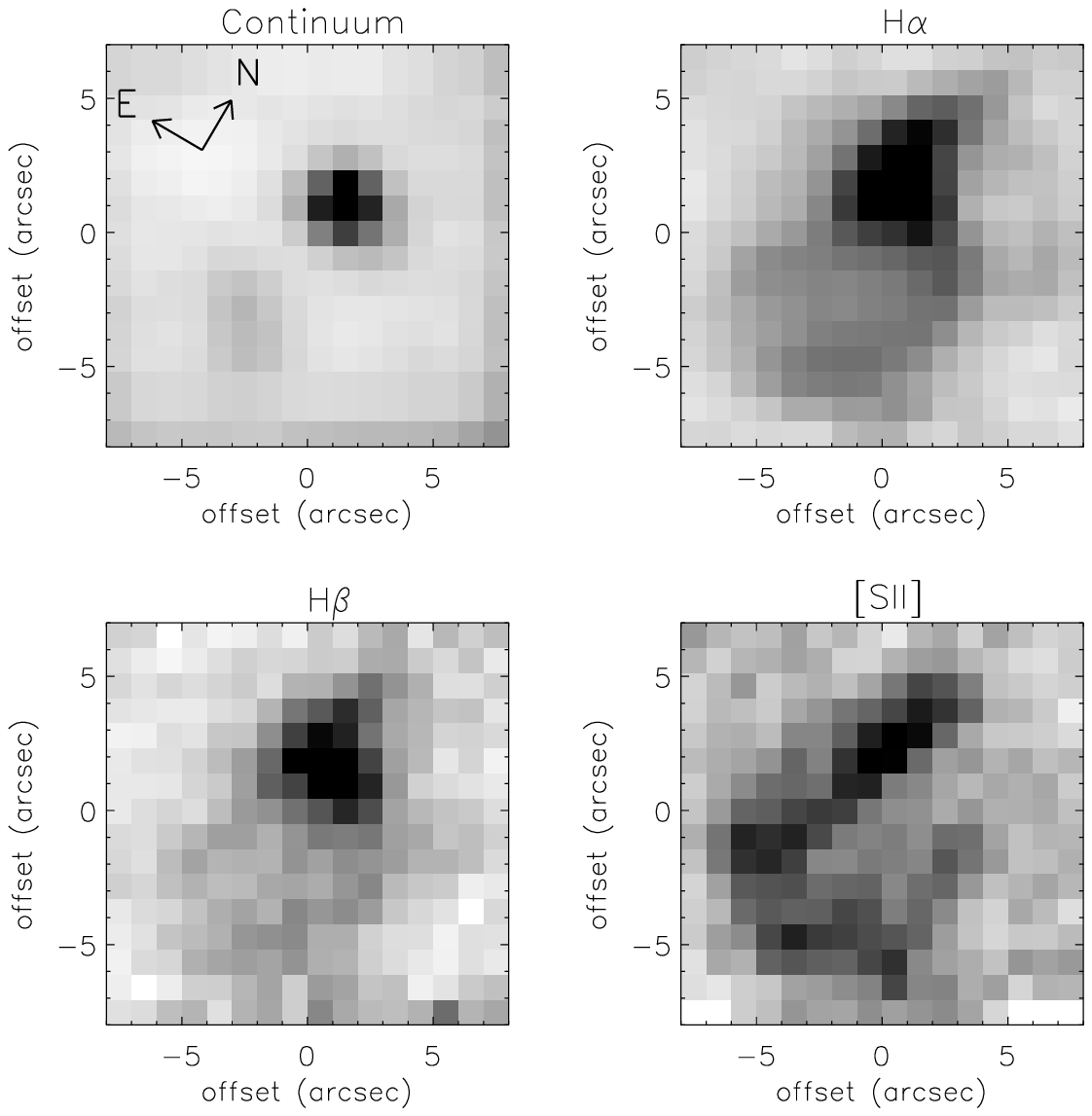}
\caption{
Left: H$\alpha$ image of the
var\,2  region taken by Massey et al. (2001) and the MPFS field with the line
izophotes superimposed. Right:
Monochromatic and continuum MPFS maps of var\,2.
A bipolar nebula is clearly seen in the H$\alpha$ and H$\beta$
emission line maps.
It is non-symmetrical and presents different morphology in lines of different
excitation. The star itself is a source of emission in permitted lines.
The bipolar nebula 20x40 pc is shock excited.
The H$\alpha$ radial velocity gradient $\pm 30$\,km/s was detected along
the bipolar structure.
}
\end{figure}

\section{Results and discussion}

Practically all known LBV stars have circumstellar nebulae
(Humphreys and Davidson~1994). Typical galactic LBV nebulae have sizes in the
range of 0.1--4~pc, expansion velocities 15--100 km/s, and their
dynamical times
are in the range 100--$5 \cdot 10^4$~years (Weis~2003).
Numerical models of the nebular expansion around massive stars
(Garcia-Segura et al., 1996) have shown that those nebulae may reach diameters
up to 20--40 pc at the main-sequence and pre-LBV stages.

Fabrika et al. (2005) have studied two LBV-like stars and their
nebulae in M\,33. The first one, B416 is a B[e]-supergiant with an expanding
ring-like nebula 20$\times$30~pc. In the second, v532 (LBV/Ofpe star) they found
an elongated nebula. The both stars' nebulae show radial velocity
gradients of about a few tens km/s. We started a special study of
gas environments around classical LBV stars in M\,33 to confirm a presence
and to study of large-scale nebulae in these objects.

We found the large-scale nebulae  around
LBV stars var\,B, var\,2, var\,83. The structure of the nebulae
indicates that they were formed by the LBV (or pre-LBV) winds.
\begin{figure}
\includegraphics[width=5.7cm]{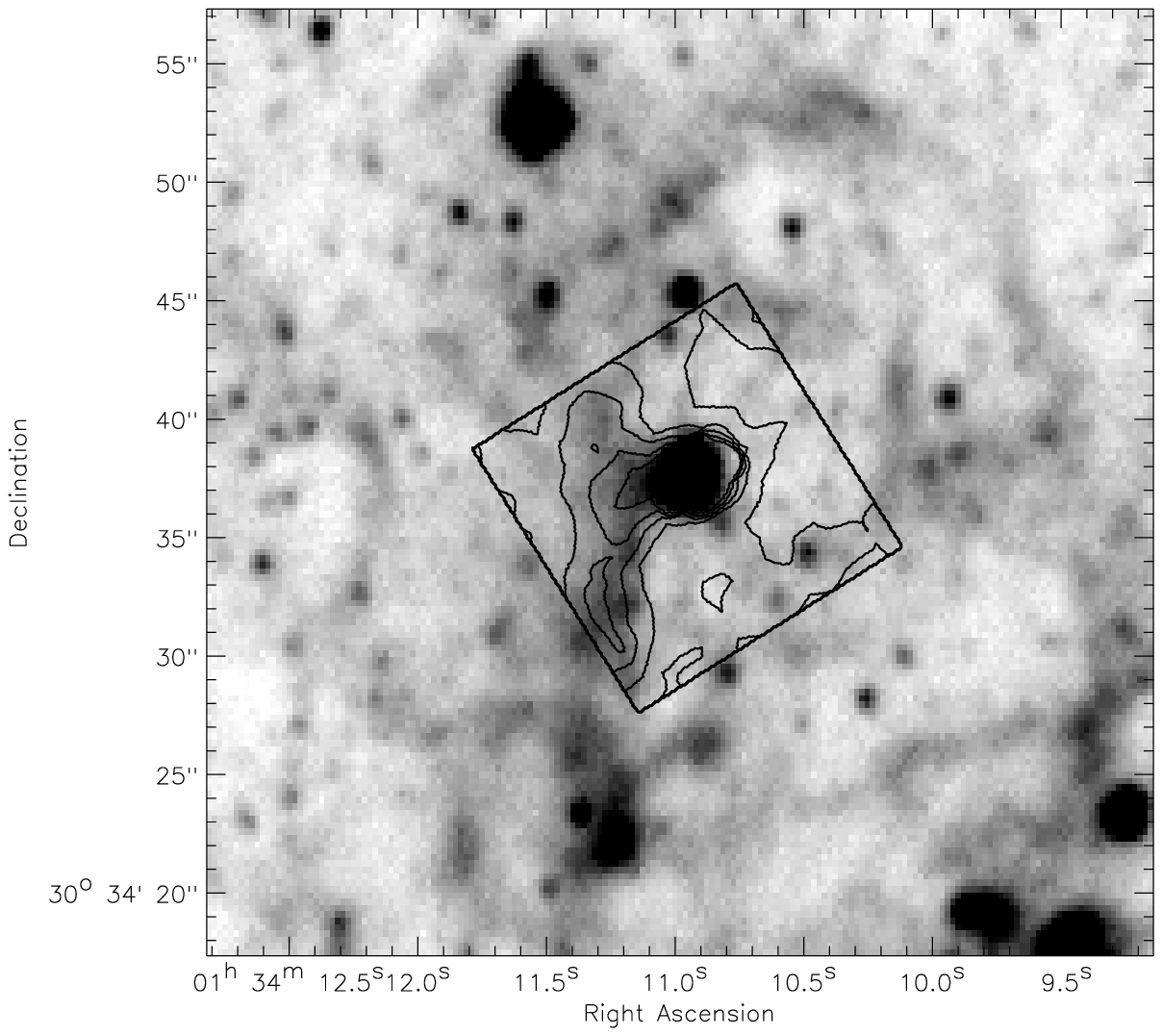}\hspace{0.5cm}
\includegraphics[width=4.8cm]{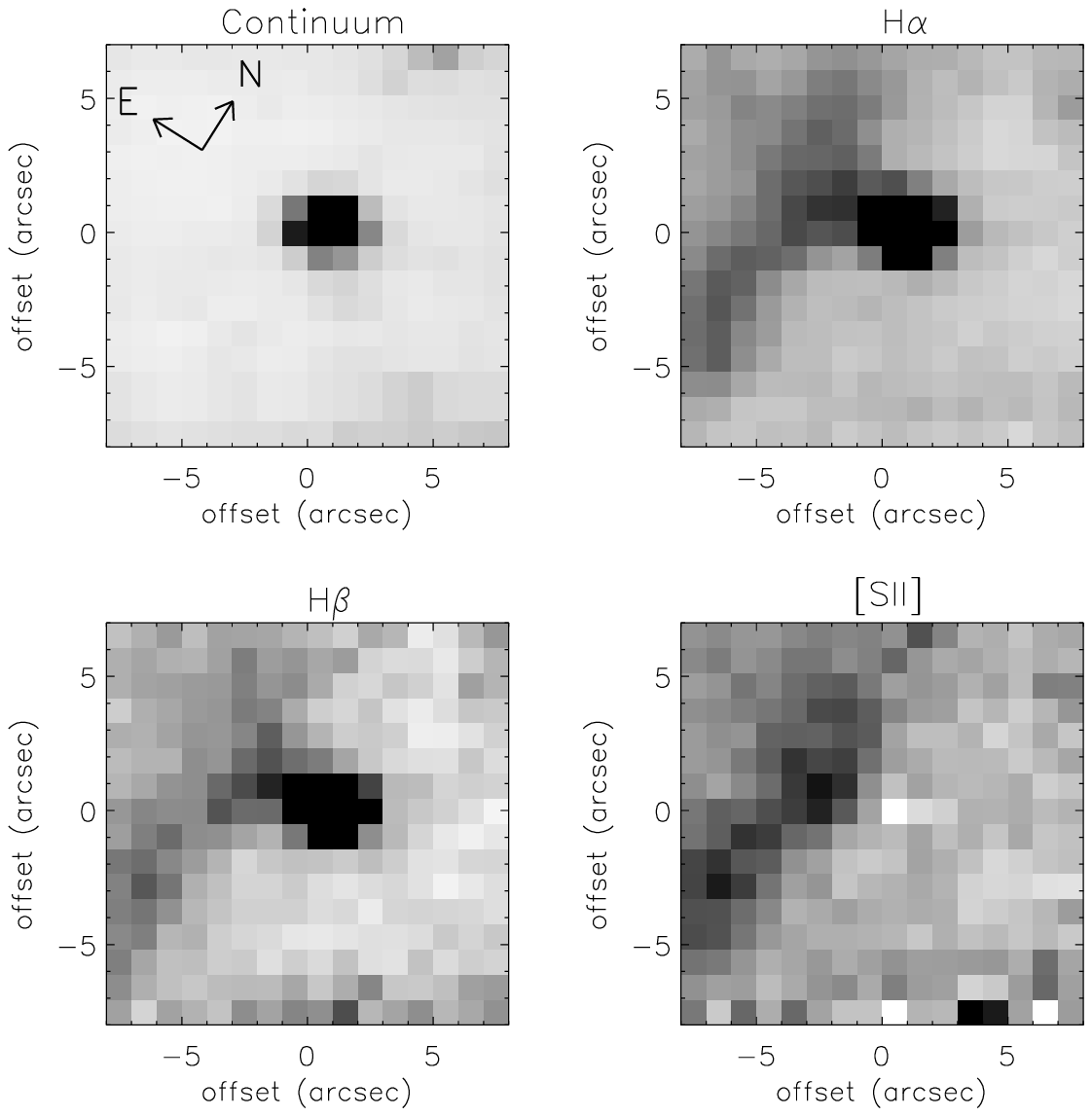}
\caption{
Same as in Fig.\,1 for var\,83.
In H$\alpha$, H$\beta$, [SII] lines we see
a double-tail-like nebula, which is a part of  a
bigger nebula. Velocity maps in H$\alpha$ show common nature of
the star and the  nebula. The double tail approaches us with a
velocity $\approx 15$~km/s.
}
\end{figure}
The nebulae are kinematically connected with the host stars.
Their physical extension is about 15~-- 30~pc,
and their dynamical times are in the range of $10^5 - 10^6$~years.
The stars  var\,A and  var\,C do not show extended nebulae,
but nebular lines are certainty present in their spectra.

\begin{figure}
\includegraphics[width=4.5cm,angle=90]{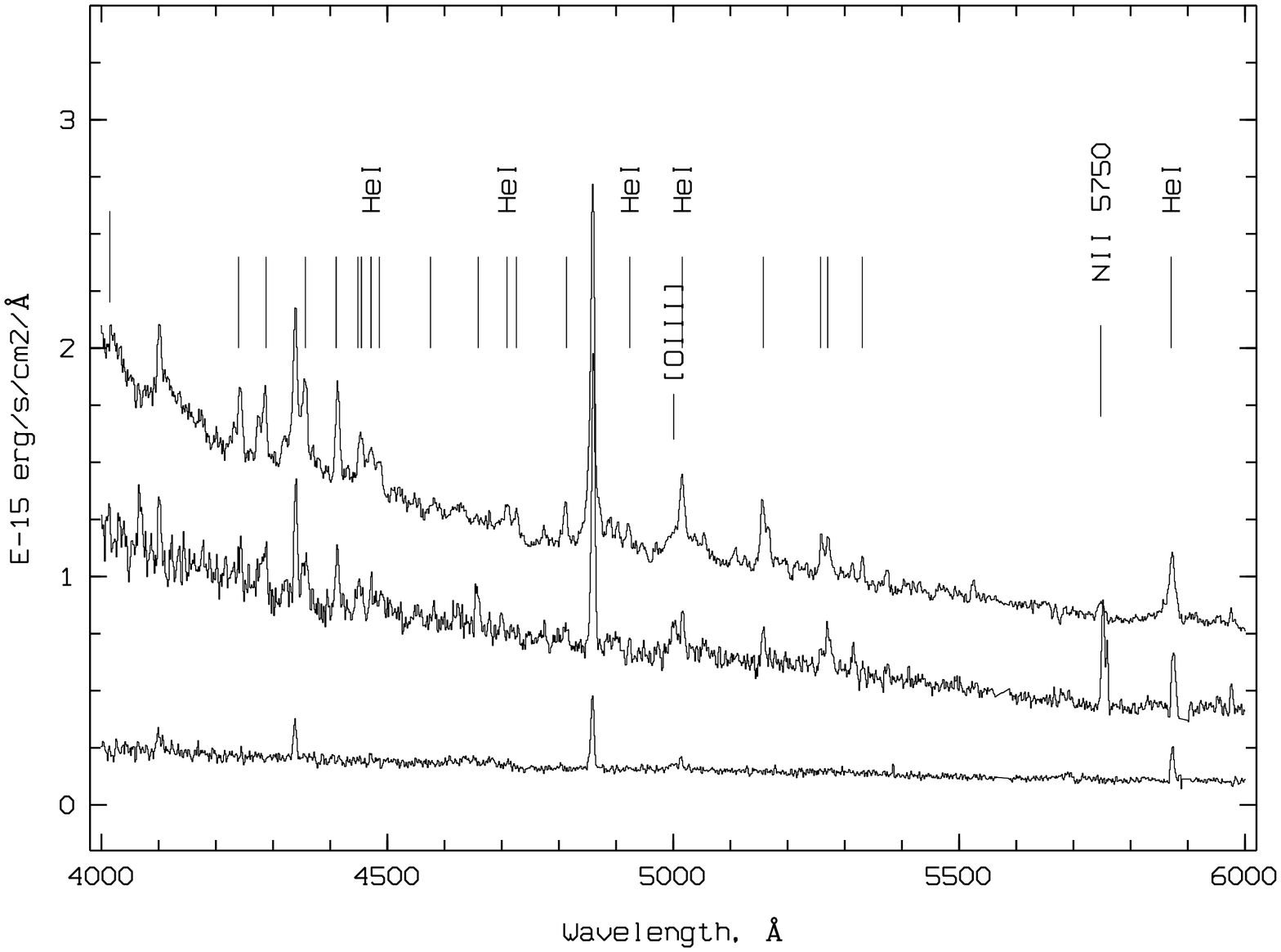}
%\hspace*{0.5cm}
%\vspace{-0.5cm}
\includegraphics[width=5cm]{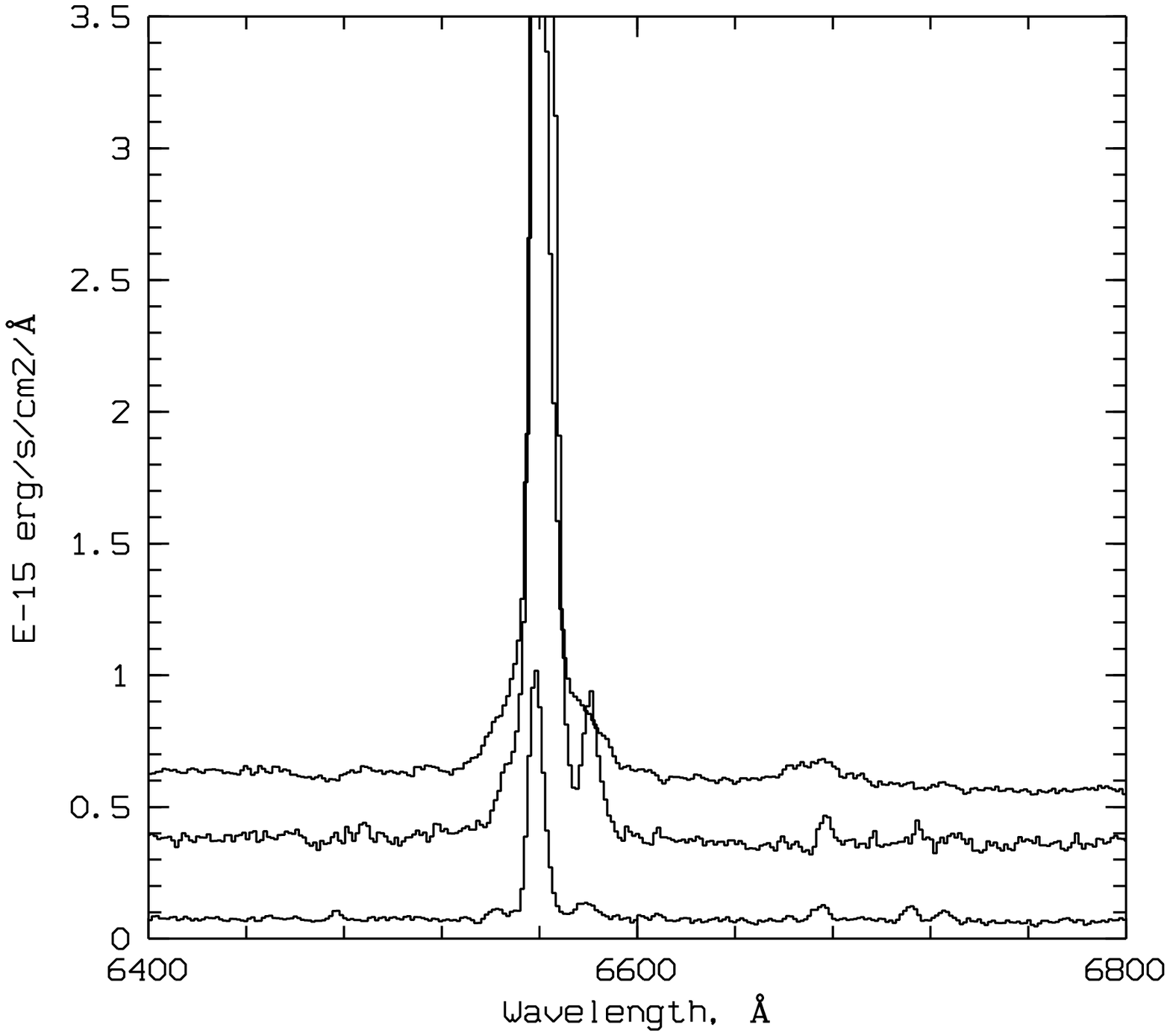}
\caption{
 Fragments of the spectra of var 2,  var B,  var 83 are shown from bottom
to top. The brightest
spectral lines are hydrogen lines and He I. Fe II, [FeII] lines shown
by unlabeled
vertical lines.
In H$\alpha$ lines we see broad components, that indicate the stellar wind.}
\end{figure}

Nebulae of such dimensions around LBV-stars in our Galaxy can not be
studied, because their diameters would exceed
1 degree. Detection of large-scale nebulae around LBV-stars
is important, a study of these nebulae can give information about the
earliest phases of evolution massive stars.

\section {Acknowledgements}
This work has been supported by the RFBR grants N\,03--02--16341,
04--02--16349 and  RFBR/JSPC grant N\, 05--02--19710. O.\,Sholukhova, P. Abolmasov, S. Fabrika
are grateful to the SOC and LOC of the
Workshop "Science Perspectives for 3D Spectroscopy" for support.

%
% BibTeX users please use
\bibliographystyle{}
\bibliography{}
%
% Non-BibTeX users please follow the syntax
% the syntax of "referenc.tex" for your own citations

\printindex
\end{document}